\documentclass[preprint]{aastex}

\begin{document}

\title{High-resolution Near-Infrared Images and Models of the 
Circumstellar Disk in HH~30}


\author{Angela S. Cotera}
\affil{Steward Observatory, University of Arizona,
    Tucson, AZ 85721}
\email{cotera@as.arizona.edu}

\author{Barbara A. Whitney}
\affil{Space Science Institute, 3100 Marine Street, Suite A353, Boulder, CO 80303}
\email{bwhitney@colorado.edu}

\author{Erick Young}
\affil{Steward Observatory, University of Arizona,
    Tucson, AZ 85721}
\email{eyoung@as.arizona.edu}

\author{Michael J. Wolff}
\affil{Space Science Institute, 3100 Marine Street, Suite A353, Boulder, CO 80303}
\email{wolff@colorado.edu}

\author{Kenneth Wood, Matthew Povich}
\affil{Harvard-Smithsonian Center for Astrophysics, 60 Garden Street, Cambridge, MA 02138}
\email{kwood@cfa.harvard.edu}

\author{Glenn Schneider, Marcia Rieke,and  Rodger Thompson}
\affil{Department of Astronomy and Steward Observatory, 933 N. Cherry, University of Arizona, Tucson, AZ 85721}
\email{gschneider@stsci.edu,mrieke@as.arizona.edu,rthompson@as.arizona.edu}

\begin{abstract}
We present Hubble Space Telescope (HST) Near-Infrared Camera and 
Multi-object Spectrometer (NICMOS) observations of the reflection nebulosity associated 
with the T~Tauri star HH~30.  The images show the scattered light 
pattern characteristic of a highly inclined, optically thick disk with a 
prominent dustlane whose width decreases with increasing wavelength.  
The reflected nebulosity exhibits a lateral asymmetry in the upper 
lobe on the opposite side to that reported in previously published Wide Field Planetary Camera 2 (WFPC2) images.    
The radiation transfer model which most closely reproduces the data has a flared accretion disk with dust grains larger than
standard interstellar medium grains by a factor of approximately 2.1.  A single
hotspot on the stellar surface provides the necessary asymmetry to fit
the images and is consistent with previous modeling of the light curve
and images.  Photometric analysis results in an estimated extinction of 
A$_V\gtrsim$80; however, since the photometry measures only
scattered light rather than direct stellar flux, this a lower limit.
The radiative transfer models require an extinction of A$_V = 7,900$.
\end{abstract}


\section{Introduction}

Theory has long predicted that dense equatorial circumstellar accretion 
disks are formed during the collapse of low mass stars
(Cameron 1962, 1973; 
Wetherill 1980; Cassen \& Moosman 1981;
Terebey, Shu \& Cassen 1984; Adams \& Shu 1986; and references therein).
The observational evidence
for such disks has been inferred from the signature of excess 
infrared 
radiation, arising from starlight reprocessed by dust and the
liberation of accretion luminosity (Lynden-Bell \& Pringle 1974;
Lada \& Wilking 1984; Lada 1986; Kenyon \& Hartmann 1987),
and the spectroscopic signatures
of accretion (Hartmann \& Kenyon 1985, Bertout 1989 and references therein).
Radio and millimeter observations have successfully imaged the disks, 
allowing an investigation of their velocity and density structures
(\citealt{beckwith90}; \citealt*{beckwith93}; Koerner, Sargent, \& Beckwith 1993; 
\citealt*{davek95}).  Radiation 
transfer models predicted that the scattered light from disks could be imaged  
in systems where the disk is viewed close to edge-on to our line of 
sight, blocking light from the central star \citep{barb92}.  
Recent high resolution HST WFPC2 images have discovered just such 
signatures of highly inclined disks around T~Tauri stars (\citealt{burrows96}; 
\citealt{karl98}; \citealt{krist00}).

The first, and perhaps the most 
stunning, example of a disk to be detected in this manner was that of HH~30 
by Burrows et al. (1996).  Their WFPC2 images resolved the HH~30 nebulosity 
into the characteristic nebular pattern of an edge-on disk:  two reflection 
nebulae separated by a dark dustlane \citep{barb92}. Their images also show the strong jet present in HH~30, which is found to be highly 
collimated deep within the disk.  
Burrows et al. constructed scattered light models of a flaring disk for 
their HH~30 observations, estimating a disk mass of around $10^{-3}M_\odot$ 
and a scale height of 15AU at a radius of 100AU.  Their modeling of 
the scattered light images required the dust grains to have a scattering 
phase function that is more forward throwing than typical interstellar medium 
(ISM) grains.  
Wood et al. (1998) 
investigated whether protostellar envelopes could produce the scattered 
light pattern associated with HH~30, and concluded that only a very tenuous 
circumstellar envelope is required in addition to the disk.  Wood et al. 
(1998) estimated a disk mass of $2.5 \times 10^{-4}M_\odot$.  The mass estimates
of both Wood et al. and Burrows et al. are only sensitive
to small grains within the disk, and are dependent on the assumed gas-to-dust
mass ratio and dust composition.
Since the optical opacity of large grains, rocks, and 
planetesimals is small, they
have no effect on the optical and near-infrared (NIR) scattered 
light and are not well-accounted for in these mass estimates.  
If the wavelength dependence typical for ISM grains is assumed for the dust opacity, Wood et al. (1998) also 
predicted that HH~30 should appear almost pointlike in the K band (2.2\micron) 
--- a feature that we shall address in some detail in this paper.

The HST WFPC2 observations of HH~30 also showed that
the scattered light pattern is highly variable, having changed
considerably in brightness between the two observations presented by 
Burrows et al. (1996).
The variability of the scattered light nebula in HH~30 is not surprising given
that photometric variability is a ubiquitous characteristic of
T~Tauri stars, with {\it periodic} photometric variability having now been
detected in hundreds of systems
(e.g., \citealt{bouvier93};  \citealt{ch96}; \citealt{wichmann98}).  
In classical T~Tauri stars, such variability 
is attributed to hot and cool spots on the stellar surface (e.g., \citealt{bouvier93};
\citealt{kenyon94}; \citealt{herbst94}; \citealt{eaton95};
\citealt{ch96}), with magnetic
accretion models naturally providing these hot spots
(Ghosh \& Lamb 1979a,b; \citealt{konigl91}; \citealt{shu94}; 
\citealt{os95}; \citealt{najita95}).  Wood \& Whitney (1998) investigated 
the observational signatures of stellar hot spots, finding that the 
variations in the HH~30 observations by Burrows et al. (1996) could be 
explained with such a model.  They also predicted that the disk should 
brighten asymmetrically during the stellar rotation period.  Such a 
lateral asymmetry was found by Stapelfeldt et al. (1999) in another 
WFPC2 pointing at HH~30.  Wood \& Whitney modeled the asymmetry with a hotspot model 
and found that the asymmetric brightening is best reproduced with a single 
large hotspot.  More recently, Wood et al. (2000) presented ground based 
$VRI$ monitoring of HH~30 showing that it is highly variable by up to 1.5~mag 
over timescales of a few days and modeled the variability as arising from 
a single hotspot.

In order to extend these fruitful investigations of a star-plus-disk 
system to longer wavelengths, we have obtained HST NICMOS images of 
HH~30.  We present the highest resolution NIR images of this 
object to date, and the first NIR images to resolve the dustlane.  
In Section 2, the observations 
and the data reduction techniques
are discussed, and the resulting images presented. In Section 3, the 
observed morphology, the photometry, and the extinction based on the 
observations, are considered. In section 4, models 
assuming a single hot spot provides the illumination asymmetry are fit to the data, a dust opacity law which is flatter than interstellar (implying larger grains) is shown to be appropriate, and an
extinction through the disk of about A$_V = 7,900$ is dervied.  
Section 5 presents our main conclusion -- 
multiwavelength 
imagery provides a powerful tool for probing the dust 
opacity in protostellar disks.
 
\section{Observations and Data Reduction}

\begin{table}
\dummytable\label{filts}
\end{table}

\begin{table}
\dummytable\label{phot}
\end{table}

\begin{table}
\dummytable\label{params}
\end{table}

The observations were made using NICMOS Camera~2 which has a nominal 
plate scale of $0.''0755\pm0.''005$ per
pixel to obtain diffraction 
limited images with a measured FWHM of $0.''11$ at $1.104\mu$m.  
At the distance of HH~30 (140~pc, \citealt{elias78}), this corresponds to an 
angular resolution of $\sim$15~AU.
The observations were made on 29 September 1997.   Five filters 
were selected for observation, F110W, F160W, F187N, F204M, and F212N. 
Information on the effective wavelengths for these nonstandard filters, the FWHM of the filter functions, and other
relevant parameters are presented in Table \ref{filts}.
Images were taken with all five filters sequentially at each of 
four dither positions.  Total integration times for each filter are given in Table \ref{filts}. Frames used for dark subtraction were taken at the end of the observations.  

The data were reduced using the IRAF data 
reduction package NICRED written for HST/NICMOS data (\citealt{mcleod97}). Darks 
for F110W, F160W and F187N were 
created using the observed dark frames with the routine 
NICSKYDARK, part of the NICRED package. Dark frames for F204M and F212N 
were created using a few of the dark frames and the actual data frames 
using NICSKYDARK.  The flats used were those produced by M. 
Rieke (private communication) for use with the NICMOS data.  
After dark subtraction and flat fielding, the images were cleaned 
for any additional bad pixels by hand using the IRAF routine IMEDIT. 

The NICMOS pixels are non-square by $\sim$1\%, and prior to shifting and 
adding, the pixels were rectified using IDL procedures developed for
the Image Display Paradigm \#3 (IDP3) software package at Steward
Observatory, specifically for use 
with the NICMOS data \citep{stobie99}.  IDP3 allows for interactive registration and
quantitative analysis, such as polygon photometry, of images using IDL widgets. \footnote{IDP3 is publicly available at http://nicmos.as.arizona.edu/software/idl-tools/toollist.cgi}  After rectification,
additional routines developed at Steward Observatory, 
NICMOSAIC and NICSTIKUM were use to shift and 
add the images respectively. NICMOSAIC performs the
shifts by aligning them using the world coordinate system (WCS)
values, and then allows
for interactive adjustments to better align the images if necessary.  HST
pointing was found to be sufficient for these observations and no 
alignment other than the WCS coordinates was used. The routine 
NICSTIKUM was used to coadd the data using the shifts determined
by NICMOSAIC.  NICSTIKUM produced images with higher signal-to-noise 
than the IRAF routine IMCOMBINE, and were much smoother along the edges .  The 
combined images were then flux calibrated using the latest available 
values derived by M. Rieke for use with NICMOS data (private communication, see Table \ref{filts}).  
A three color composite image using the data from the wide and medium
filters is presented in Figure \ref{3color}. In order to compare these observations with previous WFPC2 results we also present a three color composite image which 
combines the WFPC2 F675W taken in December 1998 (Stapelfeldt, private communication), with the NICMOS F110W and F204M images taken in September 1997 (Figure \ref{wfpc}). 
Given the time variability of the object \citep{karl99}, the opposing asymmetries are not unexpected.

As no narrow band continuum filters adjacent to the F212N and F187N filter
observations were taken due to time constraints, a method had to be 
developed to provide a best, empirical, estimate of the continuum in 
the narrow band filters using the F204M and F160W filter for the continuum.  
At first only the F204M filter was used, but since HH~30 varies considerably with wavelength, the results for the F187N filter were unsatisfactory.  A combination 
of the F160W and F204M images was also tried, but since we do not know the intrinsic
colors of HH~30, we could not derive a credible method for weighting the filters.  
In the end, a semi-empirical method was chosen.  The filter closest in wavelength 
to the narrow band observation was selected as the continuum image.  This \lq\lq~continuum\rq\rq\ image, was then subtracted from the narrow band image interactively using IDP3.  Using both the region around HH~30 and a serendipitous star within the frame
for statistical calculations, the continuum image was 
incrementally scaled until the statistical variance in the subtracted image was minimized.
After subtraction, the standard deviation off source, $\sigma$, was 4.7$\times10^{-4}$ mJy/pixel at 1.87 \micron, and 3.3$\times10^{-4}$ mJy/pixel at 2.12 \micron. The peak flux on source was 4.7$\times10^{-3}$ mJy at 1.87 \micron, and 5.5$\times10^{-3}$ mJy/pixel at 2.12 \micron. The resulting images are presented in Figures \ref{h2image} and 
\ref{paimage}. 

\section{Observational Results}

Figure \ref{3color} shows that, although HH~30 appears
roughly symmetric, there are in fact asymmetries around both the 
axis of the dust midplane (due to the inclination of the disk) and 
the jet axis (similar to the lateral asymmetry reported by Stapelfeldt 
et al. 1999).   
These observed
asymmetries are wavelength dependent~--~the predicted 
decrease in the width of the dustlane with increasing wavelength is
seen, but there is also a shift in the peak of the scattered light with wavelength.  
To both quantify this asymmetry and to insure that accurate 
three color photometry could be performed, the FWHM of the F110W 
and F160W images were convolved to match the FWHM of the F204M 
image using the IRAF package GAUSS.  The serendipitous stellar 
point source in the frames was used to 
insure the convolved PSFs were correctly matched across the
filters. In addition, 
to facilitate accurate measurements of the 
physical dimensions of both the disk and reflection nebula, the 
images were resampled in IDP3 by a factor of four using bicubic 
interpolation.  

In an effort to provide a concise discussion of the morphology and 
observational results, a schematic of the nebula is presented in
Figure \ref{map}. The dustlane, two lobes of the reflection nebula 
and jet, suggest an internally consistent coordinate system, with the
minimum of the dustlane as the zero point of the y-axis and the 
jet axis as the zero point of the x-axis. We use this coordinate system
in Figures \ref{map}--\ref{residuals}.   
The midplane of the disk, the zero point of the y-axis, has been set by  
determining the average position of the minima of emission in the
160W image (the determined average for F160W has the smallest 
uncertainties and is between the average minima of the F204M and 
F110W images).  
Since our data do not show the jet unequivocally, the origin of the
x-axis was set using the position of the minimum disk 
thickness in the F110W image (the disk in the F110W image
shows the largest variation in thickness providing the most
accurate measurements of the minimum position). The derived
position is very near that of the jet, seen best as a slight 
protrusion in the F110W image.  
Finally, the north northeast lobe is taken as the upper lobe and 
the south southwest lobe is taken as the lower lobe.  Subsequent 
discussion will use this coordinate system and nomenclature 
to discriminate between the features of interest. 

\subsection{Photometry and Extinction}

Polygon photometry was obtained for the upper and lower lobes and
is presented in Table \ref{phot}, with the regions used for
the photometry illustrated in Figure \ref{map}. To derive an
upper limit on line-of-sight extinction towards HH~30, although not
within the circumstellar disk, we assumed that
since a typical stellar type in the Taurus-Auriga region is an 
M0 star with T$_{\rm eff}$=3850 K \citep{wood98}, an M0~V model atmosphere would
provide the best estimate of the actual stellar spectrum.  As a
Class II YSO, however, HH~30 undoubtably has infrared excess and 
using an M0~V photosphere model spectrum to determine 
reference intrinsic colors will artificially 
increase the derived extinction at NIR wavelengths (\citealt{greene95}).
Therefore, in using an M0~V spectral type as a reference star when estimating 
the intrinsic colors of HH~30, we will be deriving an upper limit on the 
extinction to the observed source.  We derived the colors for
an M0~V star at the distance of HH~30 by convolving the NICMOS filters with
the appropriate 1993 Kurucz Stellar Atmosphere
Atlas model (Table \ref{phot}).\footnote{The 1993 Kurucz Stellar Atmosphere
Atlas is available from STSCI at http://www.stsci.edu/ftp/cdbs/cdbs2/grid/k93models/}
Using the color differences between our 
measurements and the selected reference M0~V star, in conjunction with
the interstellar extinction law of \citet{rl85}, spline interpolated to the
central wavelengths of the relative NICMOS filters (see Table \ref{filts}), we 
derive an extinction of A$_V=4.1\pm0.9$ along the line of sight to the region.
This extinction is consistent with other stars in the Taurus-Auriga cloud, but
we reiterate that this is an upper limit. 

Even though we do not detect the star in the 
F204M filter, we can still use the observed flux at the presumed 
location of the star (the origin of the coordinates in Figure \ref{map}) 
to estimate a lower limit on the extinction within the dustlane of HH~30. 
If we 
can determine the minimum magnitude of a star we {\it{would}} 
have been able to distinguish in our images, we can derive an 
observational lower limit on the extinction within the dustlane.
There are two important caveats to this approach: we have
no way of knowing the exact unextincted magnitude of the star, and
therefore assume a K band magnitude $\sim$9, typical for 
Class II stars in the Taurus-Auriga Cloud (\citealt{kh95}); the light we do see is all scattered light, so the
derived extinction values are a lower limit, representing a best estimate based on the available data.  

Using the TinyTim code (\citealt{krist97}), we created an artificial
point-spread-function (PSF) with parameters that match our observations
for all of the observed filters. These artificially created PSFs were
also used when convolving our radiative transfer models (see \S\ref{models}) for comparison 
with the data. Using IDP3, we added the artificial stellar PSF scaled to a 
known magnitude in the F204M filter.  If we add a PSF star with F204M 
magnitude $\sim$18.9, the star is
detectable above the background, this is, therefore, the minimum stellar 
magnitude we could have detected at the longest wavelength.  Using the
assumed intrinsic magnitude and our interpolated interstellar extinction law, 
we derive a {\it{lower limit}} on the extinction
of A$_{\rm F204M}\sim9.9$, corresponding to an 
A$_V\gtrsim$80.  This is considerably more than
the lower limit of A$_V\sim24$ derived by Burrows et. al. (1996) (and 
significantly higher than the A$_V$=30 used in their models), but is
consistent with the lower limit of A$_V\sim$80 as suggested by \citet{ksam99}.  Yet another caveat, the derived extinction assumed an ISM 
extinction law, which is unlikely to be correct given the results of our 
investigation of the dust opacity
in the disk as discussed below.  

The best estimate of the actual extinction within the dustlane 
is that derived from our radiative 
transfer models of the scattered light (see \S4), based on the assumed density 
distribution of the dust in the disk,
our viewing angle through the disk, and the opacity law of the dust.
The extinction is computed by comparing the amount of direct (unscattered)
light escaping the system at our viewing inclination to the stellar
flux emitted into this direction, that is A$_V = -2.5 \log (F_d/F_{\star}$).
This is more accurate than integrating the optical depth through the envelope
since it accounts for the emitting area of the star which sees different
paths through the disk.
Using this method, through our best fit model, we calculate an extinction 
of A$_V\sim$7,900.

\subsection{Lateral Asymmetry}

As stated earlier, Stapelfeldt et al. (1999) reported a lateral asymmetry 
in the upper nebula in their 1998 WFPC2 image of HH~30.  We also find a 
lateral asymmetry in the NICMOS images, but {\it on the opposite side} to 
that reported by Stapelfeldt et al. (1999).  
The location of the peaks in the reflected emission for upper and lower lobes are plotted in Figures \ref{map} and \ref{cont_all}.  
As can be seen, the peaks move away from the origin with decreasing 
wavelength.  The peak in the F204M
filter image is actually very close to the zero point of the
x-axis for both the upper and lower lobes.  In the F110W image, 
the peak in the upper lobe is offset by $\sim0.''2$ (28 AU), and
$\sim0.''14$ (20 AU) in the lower lobe {\it{both}} in the negative
x-direction.  
The increasing peak offset with decreasing wavelength in the y-direction
is easily explained by extinction due to dust in the disk. 
Since the asymmetry in the x-direction is undoubtably related to the
scattering by the disk, the supposition that the offset is caused 
by a simple difference in extinction from the disk is not satisfactory.  
One possible explanation is a wavelength dependent variation in the 
forward throwing scattering off the grains.
Asymmetry in the x-direction has been noted previously 
(Burrows et al. 1996), and time variable asymmetry has been both 
predicted and observed (Wood \& Whitney 1998; Stapelfeldt et al. 1999).  
No indication of a wavelength dependence to this asymmetry, however, 
has been previously noted.  

The peak of the scattered light in the WFPC2 F675W images moves from a 
negative x-axis position in 1994 to a positive x-axis position in 1998 
(Stapelfeldt et al. 1999).  
The NICMOS observations were taken in late September 1997.  
The timescale of the variations in
the asymmetry has not been established, although the recent $VRI$ monitoring 
of Wood et al. (2000) shows large variability on timescales of a few days.  
High resolution synoptic imaging is required to test whether this 
unresolved photometric variability is related to the variability in the 
image morphology.

\subsection{Disk Morphology}

The location of the minimum emission in the disk, the maximum 
emission in the two nebular reflection lobes, and the total distance between the 
maxima have been determined for the data and the models and 
are presented in Figure \ref{disk_all}. 
In Figure \ref{disk_all}, the 
F204M data graph shows a jump in the location of reflection
nebulosity maxima at 
$-0.''2$ for the lower lobe. 
There are, in fact, two peaks in the F204M emission in this region. This 
is illustrated more fully in Figure \ref{linecut}, where the two 
peaks in the linecut at position A 
are clearly illustrated (see Figure \ref{map} for linecut positions).  This double 
peak is seen only in the F204M data. Figure \ref{disk_all} also illustrates that
the rate of change of the disk thickness with distance from the assumed
origin is asymmetric about {\it{both}} axes in the data.  We 
note that the derived disk thickness varies by wavelength as expected,  with a 
minimum difference betweed the maxima of the upper and lower lobes of $\sim$92 AU 
at F110W, $\sim$81 AU at F160W, and $\sim$74 AU for F204M.

Scattered light models predict the width of the dustlane should decrease 
with increasing wavelength (Wood et al. 1998) and thus multiwavelength 
imaging can be used to probe the disk mass and opacity.  
In Figure \ref{linecut},
the decrease in the width of the dustlane towards longer wavelengths 
is evident 
and the lateral asymmetry is clearly seen: the cut at position A has 
a significantly higher intensity than the cut 
at position C.  These vertical cuts also show that the upper and lower nebula 
are separated at all NICMOS wavelengths.  This observation conflicts 
with the prediction in Wood et al. (1998) that the F204M band image should 
appear pointlike, i.e., a vertical intensity cut should show only a single 
peak and no dustlane.  The scattered light model of Wood et al. (1998) 
assumed a wavelength dependent dust opacity that followed that of ISM 
grains --- {\it the new multiwavelength NICMOS images show that 
circumstellar dust in HH~30 must have a shallower opacity than ISM dust}.  A shallow
wavelength dependent opacity implies that larger grains sizes are needed
(see, e.g., Hansen \& Travis 1974).
In \S4, we present new models for HH~30 that exploit this multiwavelength 
observational diagnostic to determine the circumstellar dust opacity.
   
\subsection{H$_2$ and P$\alpha$ Emission}
The continuum subtracted F212N image (Figure \ref{h2image}), shows
emission at the peak of the upper lobe of the reflection nebulosity,
and emission near the presumed location
of the jet.  At the peak, the 
total flux in a polygon of $\sim$0.5 square arcseconds is 0.13$\pm$0.01 mJy.  
In the image there also appears to be two ``blobs'' along the axis of the jet, 
with the first somewhat elongated in the jet direction.  The first blob has a flux level of $\gtrsim5\sigma$, with sigma being the statistical uncertainty based on the sky value, equal to the uncertainty quoted for the region above.   The other more spherically symmetric blob a little further north has a $\gtrsim3\sigma$ flux level. The quoted values represent the largest derived uncertainty values, and are 
consequently conservative.       
We therefore believe the H$_2$ emission we are seeing is real, and may
represent a detection of the jet within the region usually dominated by
reflection nebulosity.

The continuum subtracted F187N image (Figure \ref{paimage}),  also shows
emission at the peak of the upper lobe, but nothing at the location of the
jet (which bolsters our confidence in the detection of the jet in H$_2$).  
In a polygon of $\sim$0.4 square arcseconds, we derive a flux of 0.13$\pm$0.02 mJy.
In the WFPC2 H$\alpha$ image, the jet is clearly seen (Stapelfeldt, private communication), and the lack of jet emission at Pa$\alpha$ is somewhat surprising.

\section{Models\label{models}}

Previous modeling of optical images of the disk in HH~30 focused on determining the 
size, degree of flaring, and total mass of the scattering dust particles 
(\citealt{burrows96}, \citealt{wood98}).  The asymmetries observed in the 
WFPC2 images and more recently the ground based $VRI$ variability have 
been modeled in the context of stellar hotspots that break the spherical 
symmetry of the disk illumination \citep{wood98, karl99, wood00}.  
Also suggested as explanations for the asymmetry in the images are dust
clouds near the source (\citealt{karl99}) 
or non-axisymmetric disk structure,
but detailed models have yet to confirm these suggestions.
The new NICMOS images enable us to 
further test the current disk and illumination models, and provide 
multiwavelength coverage that allows an investigation of the wavelength 
dependence of the circumstellar dust opacity.  The importance of
this diagnostic, available only with high resolution NIR data, is 
vividly illustrated by comparing the observed NICMOS F204M image 
with the prediction of Wood et al. (1998), 
whose model image at 2.2\micron\ is almost
pointlike.
Wood et al. assumed the dust opacity was one typical for ISM grains,
which decreases by a factor of $> 7$ from the R band (6500 \AA) to
the K band (2.2 $\mu$m).
The fact that the NICMOS images show 
a pattern characteristic of an optically thick disk demonstrates that 
the assumption of ISM dust grains is insufficient for HH~30---a
flatter opacity law is required to simultaneously fit both the optical and NIR
images.  

In our modeling of the NICMOS images, we not only investigate different disk 
structures and illuminations 
but also alter the wavelength dependence of the dust opacity until we obtain 
a best fit to all the images.  The scattered light models have an image 
resolution of 10.57 AU per pixel, corresponding to the pixel scale 
of NICMOS at the distance of the nebula.  We convolve each image with the 
relevant NICMOS 
TinyTim PSF to allow a direct comparison between the models and data.

\subsection{Disk and Stellar Structure}

The flared disk geometry we adopt for our modeling is (\citealt{ss73}), 
\begin{equation}
\rho=\rho_0 \exp{ -{1\over 2} [z/h(\varpi )]^2  } / \varpi^\alpha 
\; ,
\end{equation}
where $\varpi$ is the radial coordinate in the disk midplane and the 
scale height increases with radius,
\begin{equation}
h=h_0\left ( {\varpi /{R_\star}} \right )^\beta.  
\end{equation}
Recent detailed models of disk structure
find that the scale height exponent varies with radius in the
inner regions but follows a similar power law beyond a few AU (\citealt{dalessio99},
\citealt{chiang97}, \citealt{B97}).  
The scattered light models 
of Burrows et al. (1996) and Wood et al. (1998) demonstrated that high-resolution
imaging enables the determination of the disk scale height at large
radii.   In particular, they noted that the scale height
at a radius of 100 AU is $h_{100} \approx 15$~AU.
Though the scale height, $h$, is constrained, the scattered-light
models cannot discriminate 
between different combinations of $h_0$ and $\beta$ that give the appropriate $h$.

We consider three different 
disk structures from the literature, each stipulating a different value for the 
flaring parameter $\beta$: 
$\beta=9/8$ (\citealt{kh87}), $\beta=5/4$ 
(\citealt{dalessio99}), and $\beta = 58/45$ (\citealt{chiang97}).  Our
initial values of $h_0$ were chosen such that $h_{100}=15$~AU, and were
then varied slightly to get the best fitting image.  
The main differences between the resulting models were the amount
of flux intercepted and scattered by the disk, with the highly flaring 
$\beta=58/45$  disk scattering 2.7 times more than the $\beta=9/8$
disk.  We show the results for the $\beta=58/45$ disk here, but note that
all three flaring parameters give similar images as long as $h_0$ is
changed so $h_{100}$ remains constant.
We set $\alpha = 3 (\beta-0.5)$, the value appropriate for viscous accretion
theory (\citealt{ss73}).  
The circumstellar disk mass and opacity are variable parameters, with
the required mass inversely proportional to the opacity.  
The resulting best fit masses for the three 
flaring parameters were within 10\% of each other.  

In brief, the disk parameters were chosen based on theoretical models
for hydrostatic structure in disks.  The dust is assumed to be well mixed
with the gas.  The flaring exponent $\beta$ is uncertain, but the combination
of $h_0$ and $\beta$ are well-determined, giving $h = 15$ AU at a radius of
100 AU.
The density exponent $\alpha$ is chosen based on theoretical models and 
the combination of $\alpha$ and $\beta$ does not significantly change 
the resulting mass required to fit the images.  Our conclusion is that the disk
structures assumed here are reasonable approximations to the structure
observed in the HH~30 disk at radii greater than several AU.  
In the inner region, it is possible that non-axisymmetric disk structures,
dust clouds, and stellar hotspots could be contributing to the non-axisymmetric
nature of the illumination.  We consider only hot spots in this paper.

Figure~\ref{model} presents a composite three color image of our best model 
for the NICMOS images.  This model 
adopts the single hotspot illumination as described in \citet{wood00}, 
with spot temperature $T_s=10^4$~K, stellar temperature $T_\star = 3000$~K,
spot size $\theta_s = 20^\circ$, and spot latitude $\phi_s = 60^\circ$.    
For disk parameters in this model we use
a disk scale height exponent of $\beta=58/45$, density exponent
$\alpha=2.36667$,
stellar radius R$_\star$ = 2.5 R$_\odot$,   
inner radius R$_{in} = 6~$R$_*$, 
and scale height at the stellar surface of $h_0 = 0.011$.
The disk extends to 200 AU in radius.

\subsection{Dust Properties}

The models are run at wavelengths corresponding to
the effective wavelength for the F110W, F160W, and F204M filters (see Table \ref{filts}).
To insure that calculating the models at a single wavelength 
is as accurate as calculating over the entire filter function,
we ran a model finely spaced in wavelength and weighted the flux
by the NICMOS filter functions.  The result did not vary noticeably
from simply using the filter effective wavelength.

The dust opacity was modified to fit the variations in the dustlane
thickness as a function of wavelength.  Altering the disk properties~--~scale
height, flaring, radial power law~--~has no effect on the wavelength dependent width.
The dustlane thickness is related to the optical depth of the dust,
which is proportional to the opacity of the dust.  As the opacity of the dust 
decreases at longer wavelengths, the dustlane width decreases.
Our best model uses opacity ratios of  
$\kappa_{F110W}:\kappa_{F160W}:\kappa_{F204M} = 1.37:1.29:1.00$. 
This opacity
law is flatter than that for ISM grains, which 
have $\kappa_{F110W}:\kappa_{F160W}:\kappa_{F204M} = 1.86:1.58:1.00$ 
\citep{kim94}. 

To obtain this opacity law, we computed a grain size distribution
consisting of homogeneous spheres composed of either amorphous carbon
(\citealt{roul91}) or revised astronomical silicate (Weingartner 
\& Draine 2000).  The size distribution for each component is specified
using a power-law with exponential decay ($r^{-p} exp(-r/r_c)$, 
e.g., Kim, Martin, \& Hendry 1994) with a geometric cross-section ($\pi r^2$) 
weighted average radius (summed over both components) of 
$r_{eff}$~=~0.056~\micron.
The relative numbers of each grain type are such that
the ratio of C/Si is roughly solar with a gas-to-dust mass ratio of 100.
The absolute value of the opacities, and therefore the disk mass
determination, depends on the assumed value of
the gas-to-dust ratio which is not well constrained in the
circumstellar environment modeled here.
The absolute value of the opacities are also sensitive to the
exact nature of the size distribution.  Since we are only modeling
NIR wavelengths, we are effectively limiting our analysis
to grains of the order of 1~\micron\ in radius.
Thus our mass estimate is a lower limit; that is,
we can imagine a population of much larger particles
that dominate the disk mass
but contribute little to the opacity at NIR wavelengths.
What is certain in the models of the HH~30 disk
is that the opacity law is flatter than that
for interstellar grains, and this implies that the grains seen
by NIR radiation are larger than those in the interstellar
medium.
The effective grain size in this population is about 2.15 times larger
than standard ISM grains \citep{kim94}.
This is consistent with the work of Burrows et al. (1996) and
Wood et al. (2000) who found that the phase
function for the dust is more forward-throwing.

The scattering phase function is calculated using Mie theory for
spherical particles.  Tables of the scattering phase function are
read into the Monte Carlo program and used to
compute scattering directions in the radiative transfer. 
This is more precise than using the Henyey-Greenstein function for
scattering, but does not change the results significantly.
Table \ref{params} tabulates the grain properties.  The opacity $\kappa$
is given in units of cm$^2$/gm. The albedo, $a$, is the ratio of scattering cross section to absorption cross section, and thus a measure of how much light is
scattered at each interaction.  The average-cosine of the scattering
angle, g, quantifies the ``forward-throwing'' scatter from the dust; g=0 is
isotropic, g=1 is 100\% forward scattering.  The maximum polarization
obtainable for a given scatter, $P$, occurs at about a 90\arcdeg\
scattering angle.

Figure \ref{cont_all} presents contour maps and the derived emission peaks 
for the data, the modified dust opacity model (best
fit) and the models that use an ISM 
opacity law. The ISM model used here differs from Wood et al. (1998) in 
that the disk mass is set at the wavelength of the F110W filter, 1.14 $\mu$m, 
instead of the R-band filter, 6500 \AA.  The disk mass required
to fit the F110W image with ISM dust is $1.1\times 10^{-3}M_\odot$, roughly
four times larger than that required to fit the R-band image (Wood et al. 1998).
Using this mass and ISM grains in an R band model 
gives an image with too wide a dustlane, providing additional verification 
of our primary conclusion:  the ISM dust opacity law is too steep to simultaneously
fit the multiwavelength imaging data of HH 30.
Figures \ref{cont_all} and \ref{disk_all} illustrate that even when using this 
increased disk mass with an ISM grain model, the longer-wavelength
F204M model image does not match the
data image since no dustlane is produced.  
Figure~\ref{disk_all} also illustrates how changing the circumstellar dust 
opacity as a function of wavelength determines the accuracy with which the 
model reproduces the width of the dustlane at the various wavelengths.

The shallow opacity law required to match the image morphology indicates 
the presence of larger-than-ISM grains, consistent with the evidence for 
grain growth in circumstellar environments found by 
other research groups (e.g. \citealt{burrows96}; \citealt{lr98}).  
Dust is expected to settle to the disk midplane, leaving only the smallest 
grains in the upper disk layers where most of the observed radiation scatters 
(see Ruden 2000, and references therein).  Our result that a  
flatter opacity law is required
suggests that coagulation has occurred, since even the smallest grains
in the upper layers are larger than standard ISM.  

The circumstellar dust mass for our model is 
$M_{dust}=6.7\times 10^{-4}M_\odot$, intermediate between the 
the Burrows et al. (1996) and Wood et al. (1998) estimate.
Again, we emphasize that we are estimating 
the mass of only those particles that efficiently scatter light
at these wavelengths (the small grains) since larger particles
are effectively invisible.  
Additional observations at millimeter and submillimeter wavelengths 
are needed to probe the mass of larger particles in the disk. 
In addition, our mass estimate depends on 
several uncertain parameters, such as 
the composition of the grains, and the mass gas-to-dust ratio.

\subsection{Lateral Asymmetry}

The inclusion in the models of a hot spot reproduces the gross 
characteristics of the lateral asymmetry of the upper lobe (Figure \ref{cont_all}). 
The wavelength dependent shifts in the emission peaks are generally reproduced
with the modified dust opacity models, and noticeably less so in the ISM grain models. In the ISM grain model, for the lower emission lobe, no peak is produced
for F204M, while two are produced for F110W, with the stronger peak on the
opposite side from the data peak.

To investigate the more subtle differences between the models and data, 
Figure~\ref{residuals} shows the residual images formed by subtracting 
the models from the data.  These images
were produced by minimizing the statistical variance in the subtracted images using the same method as that used for continuum subtraction of the narrow band images:
first the models were scaled to the data, then the alignment
between the two images was adjusted until a minimum was obtained.  
The flux in these residual images represents $\lesssim$3\% of the 
{\it{total}} flux in the F110W and F160W NICMOS images, and $\sim$10\% in F204M, 
demonstrating how well the models fit the data. 
In the localized regions where the discripancies are more apparent, the residuals are generally $\sim$2-4\% larger.   

A careful examination of Figure \ref{residuals} produces 
a few noteworthy differences between the data and models.  
In both the F110W and F204M residual images there is an enhancement 
along the central y-axis, arguably caused by emission from the jet which is not included in the models.    The models are consistently wider and flatter in the lower nebula and stronger in the middle of the upper nebula than the data, most likely the result of asymmetries in the disk which were are also not included in the models.  Overall, the models provide an 
excellent match to the observations and these subtle variations are not 
surprising considering we have adopted an idealized, single hotspot and axisymmetric disk model. 

\subsection{Properties of the Central Source}

If the central star is a normal T Tauri star, the scale factor used to fit 
the model flux to the data can also be used to derive an approximation of 
the intrinsic flux of the illuminating source.  We assume the disk geometry and scattering
properties of the dust are correct and estimate the intrinsic flux of the 
source in the different filters; the results are presented in Table \ref{phot}.
The error bars given for the F204M magnitude are due to the uncertainty in
the flaring exponent of the disk; one magnitude spans the range of scattered
flux given by reasonable flaring exponents (from 9/8 to 58/45).  
The error bars given for the colors are due to uncertainties in the 
reddening through the Taurus molecular cloud to the HH30 disk (\S3.1).
The intrinsic colors derived for HH 30 are consistent with both normal T Tauri stars
\citep{hk85},  and T Tauri stars undergoing magnetic accretion
\citep{kenyon94}.
The intrinsic fluxes can be used to estimate the luminosity of the central
star.  Assuming a spectral type for the central source of M0 (Kenyon et al. 1998),
a distance to the source of 140 pc, and colors and bolometric correction
for an M0 source (Kenyon \& Hartmann 1995), we calculate that the
luminosity lies in the range
$L_{bol} \sim 0.2-0.9 L_\odot$.  Our calculation was performed at the
F110W wavelength (where emission from the disk is minimized), for a range
of flaring parameters, $\beta$ = 9/8 to 58/45, and assumed foreground extinction values, A$_V = 2$ to 4: hence, the range
in luminosity.

\section{Summary and Discussion}

We have presented HST NICMOS observations and detailed models of the scattered light disk in HH~30.  The NICMOS images show the characteristic 
scattered light pattern of an optically thick, highly inclined circumstellar 
disk.  There is a lateral asymmetry of the upper disk lobe which is on 
the opposite side to that seen in the 1998 WFPC2 images of Stapelfeldt 
et al. (1998).  Modeling how the width of the dustlane changes with 
wavelength has enabled us to determine the NIR wavelength dependence of 
the circumstellar dust opacity.  We find that the opacity variation with
wavelength is more shallow than that of typical ISM grains.  Thus, 
multiwavelength imaging provides a 
crucial diagnostic of circumstellar dust.  

Our estimate for the mass of the circumstellar dust
disk is $M_{disk}= 6.7\times 10^{-4}M_\odot$, but 
we caution that this represents a minimum mass, since NIR observations 
are responsive only to small particles.  The size
distribution of large particles (i.e., rocks) is not constrained
at these wavelengths (see \S4.2), and would contribute little to 
our derived opacity (D'Alessio, Calvet, \& Hartmann, 2001).
A more massive disk would also result in a larger disk accretion rate
(D'Alessio et al. 1999), and may be more consistent with the
powerful jet observed in HH 30.

Comparison of the model fluxes to the data can be used to estimate
the intrinsic T Tauri star flux and luminosity.  The derived colors 
and fluxes (Table \ref{phot}) are reasonable for a T Tauri 
star, and suggest a luminosity of 
approximately 0.2 - 0.9 $L_\odot$.
The model estimates $A_V\sim7,900$ through the disk: significantly
larger than that derived from the photometry which measures only scattered 
light from the disk, not direct stellar flux.

We have modeled the lateral asymmetry of the upper reflection nebular 
with a single hotspot on the stellar surface.  
This model is also consistent with the observed optical
variability of 1.5~mag, seen over a period of a few days \citep{wood00}.
The asymmetry is variable, with the nebular 
shape appearing different in each epoch of the published WFPC2 data 
(\citealt{burrows96}; \citealt{karl99}), and in 
our NICMOS images.  The timescale of the morphological variability is as yet 
undetermined.   High resolution synoptic imagery is required to determine 
if the nebular variability is correlated with the photometric variability.  
The short timescale variability suggests changes occur very close to the 
star. One of the major goals in future monitoring of HH~30 
should be to resolve whether this variability is caused by stellar hotspots, 
as presented above, or by shadowing of warps in the inner disk 
(\citealt{karl99}).
 
\acknowledgements
We would like to thank NASA for both the telescope time and funding, without which none of the research for this paper would have been possible. 
AC, EY, GS, MR \& RT acknowledge support under NAG5-7923. 
We also acknowledge financial support from NASA's Long Term Space Astrophysics 
Research Program, NAG5~6039 (KW), NAG5~8412 (BW); the National Science 
Foundation, AST~9909966 (BW and KW); the HST Archival Research Program,  
AR-08367.01-97A, (BW, KW).  We thank Karl Stapelfeldt for
providing the WFPC2 F675W image and for helpful comments. We also thank the anonymous referee for comments which improved the final manuscript.

\clearpage

\figcaption{Three color composite image of HH~30.  F110W is blue, F160W is green
and F204M is red. A schematic of the disk and reflection nebulae is presented
in Figure \ref{map}. \label{3color}}

\figcaption{Three color composite image of HH~30.  WFPC2 F675W is blue, NICMOS F110W
is green and NICMOS F204M is red.  This figure clearly illustrates how the 
lateral asymmetry of the nebula has changes with time.  The WFPC2 image 
was taken in December 1998; the NICMOS data were obtained in September 1997.  \label{wfpc}}

\figcaption{Inverted grayscale image of the continuum subtracted F212N, H$_2$,  
image.  The image has been overlaid with contours from the F204M image.  
The orientation of the image is that given in Figure \ref{map}. The
peak of the continuum image corresponds to H$_2$ emission, and a very 
faint trace of the jet can be discerned going up from the peak. \label{h2image}}

\figcaption{Inverted grayscale image of the continuum subtracted F187N, Pa$\alpha$, image.  The image has been overlaid with contours from the F204M image.  The orientation of the
image is that given in Figure \ref{map}. \label{paimage}}

\figcaption{Schematic of HH~30 with the coordinate axis defined by the
jet and midplane of the disk.  Solid contours correspond to the F204M
image, the dashed lines give the region used to determine the
photometry in all images.  \label{map}}

\figcaption{Contour maps of the data and models.   Two different dust grain
opacity parameters are shown, the best fit and the ISM grains. The dots indicate 
the derived peak positions.  Contours have been scaled to clearly illustrate the positions of the peaks.  \label{cont_all}}

\figcaption{{\it{Upper:}} Location of the determined maxima for the upper 
and lower nebulae and the minima in the disk for the data and models as a function positional offset from the nominal center. {\it{Lower:}} Total disk thickness by filter taken as the total difference between the maxima of the upper and lower nebulae for the data and models as a function of offset along the x-axis.
\label{disk_all}}

\figcaption{Linecuts at fixed x-axis locations across the upper and lower nebulae.
The illustrated linecuts are at the positions
indicated in Figure \ref{map}.  At A, the double
peak in the F204M filter image is clearly seen.  \label{linecut}}

\figcaption{Three color composite image of HH~30 models.  The model F110W image is blue, F160W is green and F204M is red. \label{model}}

\figcaption{Residual images derived by subtracting the relevant models from the data.  In both the F110W and F204M images an enhancement along the central y-axis is likely emission from the jet: not included in the models. The total flux in the residual images is
typically $\lesssim$5\% of the original flux. \label{residuals}}


\begin{thebibliography}{DUM}

\bibitem[Adams \& Shu (1986)]{adams86} Adams, F. C., \& Shu, F. H. 1986, \apj, 308, 836



\bibitem[Beckwith et al. (1990)]{beckwith90}  Beckwith, S. V. W., Sargent, A. I., Chini, R. S., \& Guesten, R. 1990, \aj, 99, 924

\bibitem[Beckwith \& Sargent (1993)]{beckwith93}  Beckwith, S. V. W., \& Sargent, A. I.  1993, \apj, 402, 280

\bibitem[Bell et al. (1997)]{B97}
Bell K.R., Cassen M.P., Klahr H.H., Henning Th. 1997, ApJ 486, 372.

\bibitem[Bertout (1989)]{bertout89} Bertout, C. 1989, ARAA, 27, 351.

\bibitem[Bouvier et al. (1993)]{bouvier93} Bouvier, J., Cabrit, S., Fernandez, M., Martin, E. L., \& Matthews, J. M. 1993, \aap, 272, 176  

\bibitem[Burrows et al. (1996)]{burrows96} Burrows, C. J., et. al. 1996, \apj, 473, 437

\bibitem[Cameron (1962)]{cameron62} Cameron, A.G.W. 1962, Icarus, 1, 13.

\bibitem[Cameron (1973)]{cameron73} Cameron, A.G.W. 1973, Icarus, 18, 407. 

\bibitem[Cassen \& Moosman (1981)]{cm81} Cassen, P., \& Moosman, A. 1981, 
Icarus, 48, 353.

\bibitem[Chiang \& Goldreich (1997)]{chiang97} Chiang, E. I., \& Goldreich, P. 1997, \apj, 490, 368 

\bibitem[Choi \& Herbst (1996)]{ch96} Choi, P. I., \& Herbst, W. 1996, \aj, 111, 283

\bibitem[D'Alessio et al. (1999)]{dalessio99} D'Alessio, P., Calvet, N., Hartmann, L., Lizano, S., \& Cant\'o, J. 1999, \apj, 527, 893

\bibitem[D'Alessio et al. (2001)]{dalessio01} D'Alessio, P., Calvet, N., \& Hartmann, L., \apj, in press
 
\bibitem[Eaton et al. (1995)]{eaton95} Eaton, N. L., Herbst, W., Hillenbrand, L. A. 1995, \aj, 110, 1735

 
\bibitem[Elias (1978)]{elias78} Elias, J. H. 1978, \apj, 224, 857

\bibitem[Greene \& Meyer (1995)]{greene95} Greene, T. P. \& Meyer, M. R. 1995, \apj, 450, 233
\bibitem[Ghosh \& Lamb (1979a)]{gl79a} Ghosh, P. \& Lamb, F. K. 1979a, \apj, 232, 259

\bibitem[Ghosh \& Lamb (1979b)]{gl79b} Ghosh, P. \& Lamb, F. K. 1979b, \apj, 234, 296

\bibitem[Hansen \& Travis (1974)]{ht74} Hansen, J. E., \& Travis, L.D. 1974, Space Science 
Reviews, 16, 527.

\bibitem[Hartmann \& Kenyon (1985)]{hk85} Hartmann, L., \& Kenyon, S. J. 1985,
\apj, 299, 462.

\bibitem[Herbst et al. (1994)]{herbst94} Herbst, W., Herbst, D. K., Grossman, E. J.,  Weinstein, D. 1994, \aj, 108, 1906

\bibitem[Kenyon \& Hartmann (1987)]{kh87} Kenyon, S. J. \& Hartmann, L. 1987, \apj, 323, 714

\bibitem[Kenyon et al. (1994)]{kenyon94} Kenyon, S. J., et al. 1994, \aj, 107, 2153

\bibitem[Kenyon \& Hartmann (1995)]{kh95} Kenyon, S. J. \& Hartmann, L. 1995, \apjs, 101, 117

\bibitem[Kim et al. (1994)]{kim94}Kim, Sang-Hee, Martin, P. G., \& Hendry, Paul D.
1994, \apj, 422, 164

\bibitem[Koerner et al. (1993)]{davek93} Koerner, D. W., Sargent, A. I., \& Beckwith, S. V. W. 1993, \apj, 408, 93

\bibitem[Koerner \& Sargent (1995)]{davek95} Koerner, D. W., \&  Sargent, A. I. 1995, \aj, 209, 2138

\bibitem[K\"{o}nigl (1991)]{konigl91} K\"{o}nigl, A.  1991, \apj, 370, 39

\bibitem[Krist \& Hook (1997)]{krist97} Krist, J. E., \& Hook, R. N. 1997, The Tiny Tim
User's Guide Version 4.4 (Baltimore: STScI)

\bibitem[Krist et al. (2000)]{krist00} Krist, J. E., Stapelfeldt, K. R., M\'enard, R., Padgett, D. L, \& Burrows, C. J. 2000, \apj, 538, 793

\bibitem[Lada \& Wilking (1984)]{lw84} Lada, C. J. \& Wilking, B. A. 1984, 
ApJ, 287, 610.

\bibitem[Lada (1986)]{lada86} Lada, C. J. 1986, in {\it Star Forming Regions},
ed. M. Peimbert, J. Jugaku, p. 1 (Dordrecht: Reidel).

\bibitem[Lucas \& Roche (1998)]{lr98} Lucas, P. W., \& Roche, P. F. 1998, \mnras, 299, 699

\bibitem[Lynden-Bell \& Pringle (1974)]{lp74} Lynden-Bell, D. \& Pringle, J. E.
1974, MNRAS, 168, 603.

\bibitem[McLeod (1997)]{mcleod97} McLeod, B. 1997, in 1997 HST Calibration Workshop, ed. S. Casertano et al., 281 


\bibitem[Najita (1995)]{najita95} Najita, J. 1995, RMxAC, 1, 293

\bibitem[Ostriker \& Shu (1995)]{os95} Ostriker, E., \&  Shu, F. H. 1995, \apj, 447, 813
\bibitem[Rieke \& Lebofsky (1985)]{rl85} Rieke, G. \& Lebofsky, M. 1985, 288, 618

\bibitem[Rouleau \& Martin (1991)]{roul91} Rouleau, Francois  \& Martin, P.G. 1991, \apj, 377, 526

\bibitem[Ruden (2000)]{ruden00} Ruden, S. P. 2000, in Protostars and Planets
IV, V. Mannings, A. P. Boss, S. R. Russell Eds (Tucson: University of
Arizona Press).

\bibitem[Shakura \& Sunyaev (1973)]{ss73} Shakura, N. I., \& Sunyaev, R. A. 1973, \aap, 24, 337

\bibitem[Shu et al. (1987)]{shu87} Shu, F. H., Adams, F. C., \& Lizano, S. 1987, \araa, 25, 23

\bibitem[Shu et al. (1994)]{shu94} Shu, F., Najita, J., Ostriker, E., Wilkin, F., Ruden, S., \& Lizano, S. 1994,  \apj, 429, 781 



\bibitem[Stapelfeldt et al. (1998)]{karl98} Stapelfeldt, K. R., Krist, J. E., M\'enard, F., Bouvier, J., Padgett, D. L., \& Burrows, C. J. 1998, \apj, 502, 65

\bibitem[Stapelfeldt \& Moneti (1999)]{ksam99} Stapelfeldt, K. R., \& Moneti, A.  1999 in ``The Universe as Seen by ISO'', Eds. P. Cox \& M. F. Kessler, ESA-SP 427, 521

\bibitem[Stapelfeldt et al. (1999)]{karl99} Stapelfeldt, K. R., et. al. 1999, \apj, 516,  L95


\bibitem[Stobie et al. (1999)]{stobie99} Stobie, E., Lytle, D., Barg, I., \& A. Ferro, 1999, in proceedings of the NICMOS and the VLT Workshop held in Pula, Sardinia, Italy May 26-27 of 1998.  Editors Freudling, W. \& Hook, R., 77 


\bibitem[Terebey, Shu, \& Cassen (1984)]{tsc84} Terebey, S., Shu, F. H., \&
Cassen, P. 1984, \apj, 286, 529.


\bibitem[Weingartner \& Draine (2000)]{wg00} Weingartner, J.C., \& Draine, B.T. 2000, ApJ submitted, astro-ph/9907251.

\bibitem[Wetherill (1980)]{wetherill80} Wetherill, G. W. 1980, ARAA, 18, 77.

\bibitem[Wichmann et al. (1998)]{wichmann98} Wichmann, R., Bouvier, J., Allain, S., \& Krautter, J. 1998, \aap, 330, 521

\bibitem[Whitney \& Hartmann (1992)]{barb92} Whitney, B. A., \& Hartmann, L. 1992, \apj, 395, 529

\bibitem[Wood et al. (1998)]{wood98} Wood, K., Kenyon, S. J., Whitney, B. \& Turnbull, M. 1998, \apj,  497, 404

\bibitem[Wood \& Whitney (1998)]{ww98} Wood, K. \& Whitney, B. 1998, \apjl, 506, 43

\bibitem[Wood et al. (2000)]{wood00} Wood, K., Wolk, S., Stanek, K. Z., Leussis, L.,  Stassun, K., Wolff, M., \& Whitney, B. 2000, \apjl, 542, 21.
 
\end{thebibliography}
\end{document}